\documentclass{statsoc}

\usepackage[a4paper]{geometry}
\usepackage[draft]{pdfcomment}
\usepackage{graphicx}
\usepackage{amsmath, amssymb}
\usepackage[textwidth=8em,textsize=scriptsize]{todonotes}
\usepackage[paper=portrait,pagesize]{typearea}
\usepackage{natbib}
\usepackage{float}
\usepackage{multirow}
\usepackage{adjustbox}
\usepackage{booktabs}
\usepackage{bigints}
\usepackage{commath}
\usepackage{lscape}
\usepackage{psfrag}
\usepackage{dsfont}

\makeatletter
\patchcmd{\@makecaption}
{\parbox}
{\advance\@tempdima-\fontdimen2} 
{}{}

\newenvironment{felixhack}{\def\fps@table{t}}{\def\fps@table{tbp}}
\makeatother

\title[]{Classification ensembles for multivariate functional data with application to mouse movements in web surveys}

\author{Amanda Fern\'andez-Fontelo}
\address{Chair of Statistics, School of Business and Economics, Humboldt-Universit\"at zu Berlin, Berlin, Germany.}
\address{Departament de Matem\`atiques, Universitat Aut\`onoma de Barcelona, Barcelona, Spain.}
\author{Felix Henninger}
\address{Chair of Statistics and Data Science in Social Sciences and the Humanities, Ludwig-Maximilians-Universit\"at M\"unchen, Munich, Germany}
\address{Mannheim Centre for European Social Research, University of Mannheim,
Mannheim, Germany.}
\author{Pascal J. Kieslich}
\address{Mannheim Centre for European Social Research, University of Mannheim,
Mannheim, Germany.}
\author{Frauke Kreuter}
\address{Chair of Statistics and Data Science in Social Sciences and the Humanities, Ludwig-Maximilians-Universit\"at M\"unchen, Munich, Germany}
\address{Joint Program in Survey Methodology, University of Maryland, College Park, Maryland, USA.}
\author[Fern\'andez-Fontelo {\it et al.}]{Sonja Greven}
\address{Chair of Statistics, School of Business and Economics, Humboldt-Universit\"at zu Berlin, Berlin, Germany.}

\begin{document}

\begin{abstract}
We propose new ensemble models for multivariate functional data classification as combinations of semi-metric-based weak learners. Our models extend current semi-metric-type methods from the univariate to the multivariate case, propose new semi-metrics to compute distances between functions, and consider more flexible options for combining weak learners using stacked generalisation methods. We apply these ensemble models  to identify respondents' difficulty with survey questions, with the aim to improve survey data quality. As predictors of difficulty, we use mouse movement trajectories from the respondents' interaction with a web survey, in which several questions were manipulated to create two scenarios with different levels of difficulty. 
\end{abstract}


\section{Introduction}

Paradata describes information concerning the process of responding to a survey \citep{Couper2000}, often auxiliary data describing the interaction of a respondent with the survey instrument. For example, in a computer-assisted interview (CAI), paradata might reflect the time a participant takes to complete a question, or the number of back-ups or edits. Paradata attracted particular attention with the introduction of web surveys \citep{Couper2017}, as this mode is especially convenient for automatically and quickly collecting a large amount of data as a by-product of the respondent's engagement with the survey (e.g., computer mouse clicks and positions, changes of answers in a question). See  \citet{Callegaro2013} and \citet{Kreuter2010}. 

Paradata can help researchers and practitioners better understand and improve different survey aspects, including survey design and survey data quality \citep{Kreuter2013}: In CAI and web survey modes, for example, response times have been one of the most popular paradata types considered for predicting break-offs in web surveys and designing interventions \citep{Mittereder2021}, studying the relationship of response times and measurement errors \citep{Heerwegh2003} or detecting sources of difficulty in the survey \citep{Conrad2007,Yan2008}. However, response times are not always reliable descriptors of the entire survey process \citep[e.g., 
a long response time may not necessarily be due to the survey question but to other non-survey-related tasks such as checking or responding to emails, see][]{Horwitz2017,FernandezFontelo2021}. Web surveys, in particular, provide another promising source of paradata in the form of computer mouse movements, which have recently demonstrated the ability to convey additional information beyond response times \citep{Ohora2016,Stillman2017,Horwitz2017,Horwitz2019,FernandezFontelo2021}. 

Mouse movements have been used so far in a number of applications in different fields \citep{Freeman2018,Chen2001}. In survey research, mouse movement measures (i.e., features of mouse movement patterns such as the number of changes in direction or the number of times a participant is inactive for a certain period of time) have also shown promising results: \citet{Stieger2010}, for example, found that several mouse movement measures in an online questionnaire (e.g., longer inactivities or an excessive number of clicks) had a negative correlation to data quality, and \citet{Horwitz2017} demonstrated --in the controlled environment of a laboratory-- that mouse movement measures other than response times were good predictors of respondents' perceived difficulty with an online survey question. Recent studies in the field showed that several mouse movements reflected issues with items of a web survey \citep{Horwitz2019} and further that these measures were also good predictors of such issues \citep{FernandezFontelo2021}. Although all previous papers have demonstrated the potential of a number of mouse movement measures as scalar-valued features to improve some survey aspects, none of them have yet considered using the entire mouse movement trajectory as a bivariate function of the mouse cursor positions on the computer screen over time, which is the most detailed representation of the available information.

Therefore, we will investigate in this paper the potential of mouse movement trajectories as bivariate functional predictors of difficulty in web survey questions. Here, difficulty is of interest as it correlates with measurement errors in responses and thus worse survey data quality \citep{Yan2013}. To this end, our application is based on a web survey that contained a number of questions (henceforth ``target questions") that were experimentally manipulated to create two different levels of difficulty; for each of these target questions, mouse movements were collected \citep[more details are given in Section \ref{section:application}; see also][]{Horwitz2019,FernandezFontelo2021}. The purpose of the application is thus to find a predictive model using bivariate mouse movement trajectories as functional predictors to classify easy and difficult question settings. Applied to future surveys, this would be especially useful for survey researchers and practitioners since respondents facing difficulties are more likely to make errors when responding, but these errors may be corrected if they are identified early enough.

A number of classification methods exist that can use functional predictors. We give below an overview of the most important approaches out of a range of methods that have been proposed for classification with either functional or time series predictors (which we can view as functional predictors), and discuss, in particular, methods for multivariate functional data as relevant for our bivariate trajectories.

Following \citet{Pfisterer2019}, we can classify these methods into at least two main approaches. The first approach turns a functional classification into a non-functional one by extracting relevant features from the functions using, e.g., functional principal component analysis, wavelets, or splines \citep{Greven2017, Pfisterer2019}. Then, typical machine learning methods for classification tasks may be considered using the functional features as multivariate predictors. In the context of survey research, \citet{FernandezFontelo2021} used scalar-valued features of (bivariate) mouse movement trajectories as multivariate predictors in a number of machine learning models to predict question difficulty in a web survey successfully.

The second approach is genuinely functional, using the complete functional observations directly rather than summary features. It comprises a wide range of classification methods, which we grouped into methods proposed for (i) functional predictors and (ii) time series predictors. The first block includes (semi-)parametric methods such as generalised functional linear models \citep{Marx1999, James2002,Muller2004} and generalised functional additive models \citep{McLean2014}, which extend the classical generalised linear and generalised additive models, respectively. These models expand all model terms using appropriate basis representations (e.g., functional principal  components, wavelets, splines, etc.) and consider corresponding regularisation  \citep{Greven2017,Srivastava2016}. The model parameters can be estimated, e.g., using iteratively weighted least squares, or using component-wise gradient boosting \citep{Brockhaus2020}. The first block also includes non-parametric methods for functional data classification, which are generally more flexible because they are distribution-free and allow for modelling non-linear and non-additive relationships. Methods here include the kernel-based non-parametric curves discrimination (kNCD) \citep{FerratyVieu2003,FerratyVieu2006}, the functional $k$-nearest neighbours (fkNN) \citep{Fuchs2015}, and the kernel-based model proposed by \citet{Selk2021}. All of these methods use semi-metrics, which were first introduced by \citet{FerratyVieu2000} in the context of functional regression with real-valued responses. See also \citet{Morris2015} for further details on semi-parametric and non-parametric models for regression and classification tasks using functional predictors. Finally, the second block includes methods originally proposed for classification tasks with time series predictors, many of which can also be used with functional predictors \citep[e.g., ][]{Dempster2020,Prieto2013,Grecki2015,Mei2016,Wang2017}. See also \citet{Bagnall2016} and \citet{Abanda2018} for comprehensive reviews on univariate and multivariate time series classification methods.

Despite the large number of methods for classification using functional predictors that have already been proposed, most of them are either univariate functional methods or use feature extraction techniques. To date, little attention has been paid to the extension of univariate genuinely functional methods to the multivariate case, leaving considerable room for improvement in that regard. For example, \citet{Selk2021} proposed a non-parametric model for regression and classification tasks using continuous and categorical predictors, and univariate and multivariate functional predictors. While the model seems promising, the authors measured the functions' proximity using only the $L^2$ distance, which is sometimes too simplistic. Other examples include the methods by \citet{Prieto2013}, \citet{Grecki2015}, \citet{Mei2016} and \citet{Wang2017}, all of which allow for multivariate time series predictors, but measure time series proximity using either DTW (Dynamic Time Warping)-type distances or feature extraction methods (e.g., shapelets) to reduce the time series dimension. Although DTW is often used to measure the proximity of functions --especially if they may likely vary in speed-- there are other relevant distances (e.g., the Fr\'echet or Hausdorff distances) that are frequently used in curves comparison and carry relevant information. Ideally, a classification method should combine information based on different distance measures. To date, however, only a few of the existing methods combine weak classifiers to improve overall models' predictive performance \citep[e.g.,][]{Prieto2013,Wang2017} and, in these two cases, the weak classifiers are all built on the same distance (DTW). 

To investigate the potential of mouse movement trajectories as functional predictors of web survey difficulty --and in light of the want of genuinely functional methods capable of dealing with multivariate functional predictors-- this paper introduces new ensemble models for multivariate functional data classification as extensions of the non-parametric methods by \citet{FerratyVieu2003, FerratyVieu2006} and \citet{Fuchs2015} for univariate functional data classification. In particular, we propose two types of ensemble methods that are optimal combinations of weak learners and these weak learners are built based either on the fkNN classifier \citep{Fuchs2015} or kNCD classifier \citep{FerratyVieu2003}. Both fkNN and kNCD use semi-metrics as proximity measures between two functional observations. Semi-metrics as scalar quantities always summarise the information in these infinite-dimensional functions, and each can thus be viewed as capturing a specific aspect of the functional observations. Therefore, a challenging point is to choose the semi-metrics such that they capture the relevant functional characteristics well: \citet{FerratyVieu2003,FerratyVieu2006}, for example, constructed parametric families of semi-metrics (e.g., based on functional principal component analysis), and \citet{Fuchs2015} combined a whole set of different semi-metrics in the so-called $k-$nearest neighbour ensemble, using a linear combination of class probabilities that are estimated with the fkNN method. The linear combination weights show the importance of the corresponding semi-metrics in the ensemble. In this paper, we build on \citet{Fuchs2015}'s approach, contributing two extensions: First, while both \citet{Fuchs2015} and \citet{FerratyVieu2003} classify univariate curves, we generalise this to multivariate ($d$-variate) functional observations. The proposed collection of semi-metrics comprises both extensions of some of those in \citet{Fuchs2015} to the multivariate case (e.g., $L^p$ distance, mean, DTW, different versions of global maximums, minimums, etc.) as well as new proposals such as the Fr\'echet distance and Hausdorff distances \citep{Alt2000}, edit-type distances such as the Levenshtein and Hamming distances \citep{Navarro99}, a correlation-type distance \citep{Gabor2007}, and the Aitchison distance \citep{Filzmoser2018}, among others. We also incorporate a set of semi-metrics based on scalar or vectorial features of the functions. Second, we provide alternatives for combining predictions across weak learners into an ensemble model. A good ensemble combines the weak learners optimally and can considerably reduce the weak learners' prediction error rates. Instead of using linear combinations as in \citet{Fuchs2015}, we propose to construct our ensemble models using more flexible alternatives as super-learners in the stacked generalisation methods \citep{Wolpert1992, Breiman1995, LeblancTibshirani1996}.

The paper is organised as follows. Section \ref{sec:methods} presents our ensemble models, including different semi-metrics for multivariate functional data and several approaches for weak learners combination. Section \ref{section:application} describes our models' performance in our application, detecting difficulty in survey items based on mouse movement trajectories. Section \ref{section:discussion} closes with  a discussion.

\section{Multivariate functional ensembles for classification} \label{sec:methods}

Consider the learning sample $({\bf x}_i, y_i), \, i=1,\dots,n$, where $y_1,\dots,y_n$ are values of a categorical (not ordinal) random variable $Y$ with $L$ mutually exclusive and exhaustive classes $\mathcal{L} =\{1,\dots, L \}$. Let ${\bf x}_1, \dots, {\bf x}_n$ be realisations over $t \in \mathcal{T}  \subset \mathbb{R}$ of independent  copies ${\bf X}_1, \dots, {\bf X}_n$ of a multivariate ($d$-variate) functional random variable ${\bf X}\in \mathcal{F}$, where $\mathcal{F}$ is a suitable functional (infinite dimensional) space such as the $L^2(\mathcal{T})$ \citep{FerratyVieu2006}. Hence, ${\bf x}_i(t) \in \mathbb{R}^{d}$ for each  point $t \in \mathcal{T} $ with $d \in \mathbb{N}$; we focus here on $d \geq 2$. Where we talk about the $a$-th derivative ${\bf x}^{(a)}$ of ${\bf x}$, we assume that it exists and is in $L^p(\mathcal{T})$ for appropriate $p \in \mathbb{N}$ if required. In practice, functions are observed on a finite discrete grid in $\mathcal{T}$, where the grid may differ between observed functions and dimensions within a multivariate function. For simplicity, we focus here on the case of regularly spaced grids per univariate function and consider $t \in \mathcal{T}$ as time.

We address the following classification problem: for a new observation ${\bf x}_{*}$, infer the unknown class membership $y_{*}$ from the learning sample of multivariate functions with known class memberships. 

\subsection{Semi-metrics}\label{methods:semi}

Semi-metrics are proximity measures between general mathematical objects and can be used in particular for functional observations taking values in an infinite-dimensional space \citep{FerratyVieu2006}. We here use semi-metrics to measure distances between multivariate functional observations (e.g., between bivariate mouse trajectories). 

Formally, let $D({\bf x},{\bf x}_{*})$ be the semi-metric between the multivariate functions ${\bf x}$ and ${\bf x}_{*}$, which thus fulfils: $ D({\bf x},{\bf x}_{*}) \geq 0$, $D({\bf x},{\bf x}) =0$ and $D({\bf x},{\bf x}_{*}) \leq D({\bf x},{\bf \tilde{x}})+ D(\bf \tilde{x},{\bf x}_{*})$, $\forall {\bf x}, {\bf x}_{*}, {\bf \tilde{x}} \in \mathcal{F}$.
Although semi-metrics are similar to metrics, they differ from metrics in that $D({\bf x},{\bf x}_{*})=0 \nRightarrow {\bf x}={\bf x}_{*}$. Thus, different curves can have a ``distance" of zero, which is particularly relevant when defining semi-metrics via the functions' derivatives ${\bf x}^{(a)}$, $a \in \mathbb{N}$ (e.g., the $L^2$-distance between the first derivatives). 

We extend several semi-metrics proposed in \citet{Fuchs2015} and propose a broad collection of semi-metrics that capture different aspects of multivariate functional observations. More specifically, we propose five different families of semi-metric (cf., Table \ref{table:tab1}), which differ in whether they are computed over the raw trajectories or summaries thereof, and in whether they preserve the ordering as well as the spatial and/or temporal information in the trajectories. For other applications, this framework is extendable, allowing for the inclusion of additional user-defined semi-metrics.

The first family of semi-metrics --the lock-step family according to \citet{Mori2016}-- includes metrics that compare function values at the same time points $t$, keeping either the functions' ordering or treating them as unordered multivariate random vectors. This family includes the Manhattan ($L^1$) and Euclidean ($L^2$) metrics as well as further $L^p$ metrics for $p \geq 1$ for function comparison, and a correlation-type distance proposed by \citet{Gabor2007} for comparison of unordered multivariate random vectors \citep[see also][]{energy}. Extending the idea of the univariate correlation coefficient, \citet{Gabor2007} defined the distance correlation, $\mathcal{R}$, for two multivariate random vectors with finite first moments, satisfying $ 0 \leq \mathcal{R} \leq 1$ and $\mathcal{R}=0$ if and only if the two random vectors are independent. We  use as distance $1-\mathcal{R}$ (cf. Table \ref{table:tab1}), which is $0$ if two trajectories are perfectly correlated and $1$ if these trajectories are independent. In practice, we use the empirical distance correlation in place of $\mathcal{R}$.  

\begin{table}                                  
\caption{\label{table:tab1} Overview of the used collection of semi-metrics to measure distances between two multivariate functional observations ${\bf x} = ({x}_1, \dots, {x}_d)^T$ and ${\bf x_{*}}$, both defined over $\mathcal{T} \subset \mathbb{R}$ and $d \in \mathbb{N}$. These semi-metrics can also be computed for the corresponding $a-$th derivatives. See Supporting Material for more details on our collection of semi-metrics, including the symbol sequence semi-metric family.}
\centering
\fbox{%
\begin{tabular}{*{3}{|c}}
{\bf Family} & {\bf Semi-metric} & {\bf Definition} \\
\hline
lock-step & $L^{p}$ & $\displaystyle \left( \bigintssss_{\mathcal{T}} \sum_{k=1}^{d}\abs{x_k(t)-x_{*k}(t)}^p dt  \right)^{\frac{1}{p}}$ \\
(raw trajectories) & & \\ \cline{2-3}
&  correlation &  $1-\begin{aligned}[t] \left(\mathcal{R}^2({\bf x}, {\bf x}_*)\right)^{\frac{1}{2}}=1 - \left(\frac{\mathcal{V}^2({\bf x}, {\bf x}_*)}{\left(\mathcal{V}^2({\bf x})\mathcal{V}^2({\bf x}_*)\right)^{\frac{1}{2}}}\right)^{\frac{1}{2}}\end{aligned}$,\\
 & & if $\mathcal{V}^2({\bf x})\mathcal{V}^2({\bf x}_*) > 0$, $\mathcal{V}^2({\bf x}, {\bf x}_*)$ is the so-called \\
 & & distance covariance and $\mathcal{V}^2({\bf x})=\mathcal{V}^2({\bf x}, {\bf x})$ \\ 
\hline 
elastic & DTW &  $\underset{\gamma \in \Gamma}{\operatorname{inf}}  \Delta\left({\bf x} \circ \gamma, {\bf x}_{*}\right)$, \\
(raw trajectories) & & where $\Gamma$ is the space of all monotonically increasing, \\
&&  onto and differentiable warping functions $\gamma$ and \\
&& $\Delta(\cdot,\cdot)$ is a distance in $\mathcal{F}$, e.g, $L^2$ \\ \cline{2-3}
& Fr\'echet & $\begin{aligned}[t] \inf _{{\phi, \phi^{'}}\in \Phi} \max _{t \in\mathcal{T}} \Delta \left( {\bf x}\left(\phi\left(t\right)\right),{\bf x}_*\left(\phi^{'}\left(t\right)\right)\right) \end{aligned},$ \\
&& where $\phi,\phi^{'}: \mathcal{T} \rightarrow \mathcal{T}$ are functions in  \\
&& the space of all continuous and monotonically \\
&& increasing functions $\Phi$ and $\Delta(\cdot,\cdot)$ is a distance \\
&& such as the Euclidean in $\mathbb{R}^d$ \\ \cline{2-3} 
&  Hausdorff & $ \begin{aligned}[t] {\max}
& \left(\underset{t\in \mathcal{T}}{\max}\, \underset{t'\in \mathcal{T}}{\min} \Delta \left( {\bf x}(t),{\bf x}_*(t^{'})\right), \right. \\ & \left. \underset{t^{'}\in \mathcal{T}}{\max}\, \underset{t\in \mathcal{T}}{\min} \, \Delta \left({\bf x}_*(t^{'}),{\bf x}(t)\right)\right) \end{aligned}$ \\
&& where $\Delta(\cdot,\cdot)$ is a distance, e.g., the Euclidean in $\mathbb{R}^d$\\ \hline
svs & scalar & $\lvert T({\bf x}) - T({\bf x}_\ast)\rvert$ \\
(summaries) && for scalar summary $T({\bf x})$ of ${\bf x}$, \\
&& e.g., $T({\bf x})=\bigintssss_{\mathcal{T}} x_{k}(t) dt$ the mean, \\ && or the maximum or range for dimension $k$. 
\\  \cline{2-3}
& vector  & $\displaystyle \left(\sum_{k=1}^{r}\left(T_k({\bf x}) - T_k({\bf x}_\ast) \right)^{2}\right)^{\frac{1}{2}}$ \\
&& for vector summaries ${\bf T}({\bf x})=(T_1({\bf x}), \dots, T_r({\bf x}))^T$ of ${\bf x}$, \\
&& e.g., $T_k({\bf x}) = \frac{1}{|\mathcal{T}|} \bigintssss_{\mathcal{T}} x_{k}(t) dt$ the mean, \\
&& or the maximum or range for dimension $k$, and $r=d$. \\ \hline
composition &  Aitchison & $\displaystyle \left(\frac{1}{2r}\sum_{j=1}^{m}\sum_{s=1}^{m}\left(\ln\left(\frac{T_j({\bf x})}{T_s({\bf x})}\right)-\ln\left(\frac{T_j({\bf x}_{*})}{T_s({\bf x}_{*})}\right)\right)^2\right)^{\frac{1}{2}}$, \\
(summaries)&& where ${\bf T}({\bf x})=\left(T_1({\bf x}),\cdots,T_{m}({\bf x})\right)^{T} \in \mathbb{S}^{m}$ is a \\
&& compositional vector of $m$ parts summarising the \\
&& raw functional observation ${\bf x}$
\end{tabular}}
\end{table}

The second family of semi-metrics includes a variety of elastic-type distances, either keeping the ordering of our functional observations (DTW and Fr\'echet distances), or ignoring the ordering by treating the functions as point clouds (Hausdorff distance). Unlike semi-metrics in the lock-step family, the elastic-type semi-metrics are more flexible in that they allow for one-to-many or one-to-none mappings (i.e., the time point $t$ for one function can be matched to $t$, to another time point $t'$, to several, all remaining points or no time points of another function) \citep{Mori2016}. In particular, the DTW technique seeks a warping function $\gamma^*:\mathcal{T} \rightarrow \mathcal{T}$ in the space of all monotonically increasing, onto and differentiable warping functions $\Gamma$ over $\mathcal{T}$ that optimally aligns two (multivariate) functions to minimise a functional distance $\Delta(\cdot,\cdot)$ in $\mathcal{F}$ (e.g., the $L^2$ distance), $D_{DTW}({\bf x},{\bf x}_{*}) = \inf_{\gamma \in \Gamma} \Delta({\bf x} \circ \gamma, {\bf x}_{*})$. The trajectories may be of different lengths and vary in speed, and are in practice recorded over finite discrete grids. However, note that DTW is not a proper semi-metric since it does not satisfy the triangle inequality property. We can further consider other elastic-type distances such as the elastic distance 
based on the square-root-velocity (SRV) framework \citep{Srivastava2016,Steyer2021,Marron2021}, which is a proper metric, although we do not consider this distance in our application due to computational cost. This family also includes the Fr\'echet and Hausdorff distances \citep{Alt2000}: The Fr\'echet distance treats ${\bf x}$ and ${\bf x}_*$ as continuous functions ${\bf x}, {\bf x}_*: \mathcal{T} \rightarrow \mathbb{R}^d$. Let $\Phi$ be the set of continuous and monotonically increasing functions $\mathcal{T} \rightarrow \mathcal{T}$. The Frech\'et distance is defined as $D_F({\bf x},{\bf x}_*)=\inf _{{\phi, \phi^{'}} \in \Phi} (\max _{t \in\mathcal{T}} (\Delta({\bf x}(\phi(t)),{\bf x}_*(\phi^{'}(t)))))$, where $\Delta(\cdot,\cdot)$ is, e.g., the Euclidean distance in $\mathbb{R}^d$. The Frech\'et distance measures the largest remaining (point-wise) distance between two functions after optimal time alignment. By contrast, the one-sided Hausdorff distance from a set of points $A \in \mathbb{R}^d$ to another set of points $B \in \mathbb{R}^d$ is defined as $h(A,B)=\max_{a \in A}(\min_{b \in B}\, \Delta(a,b))$ for points $a \in A$ and $b \in B$, where $\Delta(\cdot,\cdot)$ is, e.g., the Euclidean distance. It measures the largest distance from $A$ to the nearest point in $B$. As the one-sided Hausdorff distance is asymmetric (i.e., generally $h(A,B) \neq h(B,A)$) the Hausdorff distance is defined as $D_H(A,B)=\max\left(h(A,B),h(B,A) \right)$. Application to functions requires the functions to be observed on discrete grids for $t$ (possibly of different lengths) and treats ${\bf x}$ and ${\bf x}_*$ as sets of points in $\mathbb{R}^{d}$. Semi-metrics in the lock-step and elastic-type families can be computed for observed functions or their derivatives. 

The third family of semi-metrics is the scalar-vector-summary (svs) family, which includes distances between scalar-valued or vector-valued summary features of the multivariate functional observations (or their derivatives). Examples are the Euclidean distance between the maximums, minimums or ranges of two multivariate functional observations in one of the dimensions (e.g., in our application, globMax-x (y), globMin-x (y) and globRange-x (y), respectively, for the first (second) dimension). Examples for vector-valued summaries include the Euclidean distance between the vectors of maximums, minimums or ranges in each of the $d$ dimensions (e.g., in our application, globMax, globMin and globRange, respectively). This family also includes the multivariate version of the mean distance by \citet{Fuchs2015} (i.e., the Euclidean distance between the multivariate mean vectors), as well as measure-based semi-metrics, which in our application are defined as the Euclidean distance between (potentially personalised) mouse movement measures considered by \citet{FernandezFontelo2021} and described in Section \ref{section:application}. 

The fourth compositional family of semi-metrics summarises the trajectory in a composition of time spent in a set of $m$ areas of interest (AOIs) in $\mathbb{R}^d$. It includes the Aitchison distance for compositional vectors in the simplex $\mathbb{S}^m=\{(z_1, \cdots, z_m): z_j > 0, \, j=1, \cdots, m; z_1+\cdots+z_m =1\}$. This distance between compositions uses essentially a Euclidean distance between vectors of log-ratios of all component pairs to account for the relative nature of compositions and the structure of the $\mathbb{S}^m$.

The last symbol sequence family includes a number of edit-type distances aimed at finding an optimal alignment between two sequences of symbols, here taken to indicate the location of the trajectory at time $t$ within a given AOI. Out of the edit-type distances, we consider in the following the Levenshtein and Hamming distances \citep{Navarro99}. In particular, the Levenshtein distance counts the number of substitutions (or mismatches) and insertions and deletions (or indels) required to change one sequence into another not necessarily of the same length. The Hamming distance only counts the minimum number of substitutions (mismatches) needed to convert one sequence into another of equal length. To ensure the same length across all trajectories for the Hamming distance, the sequences of symbols in our application were constructed using the time-normalised trajectories of computer screen locations. 

The families of semi-metrics can also be classified with regard to the information they maintain: In contrast to the compositional family, the symbol sequence family preserves the temporal information, and both families keep the spatial information in an aggregated fashion. The svs family extracts features of interest and thus does not keep either the spatial or temporal information in the functions. Lock-step and elastic families keep the spatial and temporal information in the functions, with the elastic family allowing for stretching and compressing of time for better matching between functions. See Table \ref{table:tab1} for a formal definition of some of the considered semi-metrics and Section SM.1 of the Supporting Material for more details on the complete collection of semi-metrics, including details of some of them in practice.

\subsection{Weak learners for multivariate functional observations} \label{sec:weakLearners}

Weak learners are the basis of ensemble building, as their combination gives a more robust model (ensemble) with potentially better predictive abilities. We focus here on two methods for weak learner construction: the fkNN used by \citet{Fuchs2015} in the context of univariate functional data classification and kNCD proposed by \citet{FerratyVieu2003} in the context of scalar-on-function regression. Both methods measure the proximity of functional observations through semi-metrics, and  we use here the collection of semi-metrics for multivariate functional data described  in Section \ref{methods:semi}

The fkNN rule predicts the class membership $y_{*}$ of the new observation ${\bf x_*}$ with the majority class among the $k$ closest neighbours of ${\bf x_*}$. We identify these neighbours by ordering the training predictors $ {\bf x}_1, \cdots , {\bf x}_n $ given a specific semi-metric $D(\cdot,\cdot)$ such that:
\begin{eqnarray}\label{eq:knnorder}
D\left({\bf x_{*}},{\bf x}_{(1)}\right)\leq \cdots \leq D\left({\bf x_{*}},{\bf x}_{(k)}\right) \leq \cdots \leq D\left({\bf x_{*}},{\bf x}_{(n)}\right),
\end{eqnarray}
where ${\bf x}_{(1)}, \cdots, {\bf x}_{(n)}$ are the ordered training predictors from the nearest to the farthest to ${\bf x}_*$ given the semi-metric $D(\cdot,\cdot)$. According to expression (\ref{eq:knnorder}), the neighbourhood $\mathcal{N}^k({\bf x_*})$ of the $k$ closest training functional predictors of ${\bf x_*}$ given $D(\cdot,\cdot)$ is defined as:
\begin{align}\label{eq:nb}
\mathcal{N}^k({\bf x_*})=\left\{{\bf x}_j: D\left({\bf x_{*}},{\bf x}_{j}\right)\leq D\left({\bf x_{*}},{\bf x}_{(k)}\right)\right\}.
\end{align}

We finally predict the class membership $y_*$ of ${\bf x}_*$ as the most frequent class of the observations in the neighbourhood $\mathcal{N}^k({\bf x_*})$. That is, $\widehat{y}_{*}=\underset{l \in \mathcal{L}}{\mathrm{argmax}}(\widehat{ p}_{l}(k))$, where $\widehat{p}_{l}(k)$ is the estimated probability of the new observation ${\bf x_{*}}$ belonging to class $l \in \mathcal{L}$ defined as:
\begin{eqnarray}\label{eq:posterior}
\widehat{p}_{l}(k)=\frac{1}{k}\sum_{{\bf x}_i \in \mathcal{N}^k({\bf x_*})} {\mathds{1}}_{y_i=l},
\end{eqnarray}
where ${\mathds 1}_{y_i=l}$ is the indicator function taking $1$ if $y_i=l$ and $0$ otherwise. Although $\widehat{p}_{l}(k)$ is dependent on the parameter $k$, this dependence will be suppressed from now on in the notation for simplicity. The estimated probability $\widehat{p}_l$ is thus the proportion of observations ${\bf x}_j$ belonging to the neighbourhood $\mathcal{N}^k({\bf x}_*)$ whose class is $y_j=l$ for $l \in \mathcal{L}$. The class $y_*$ can be predicted with the majority class among observations ${\bf x}_j \in \mathcal{N}^k({\bf x_*})$. If there are any ties in the majority class, they are broken at random. Also, all candidates are included in the vote if the ties are in the vector of the $k$ nearest neighbours. More details are found in \citet{Fuchs2015}, who focused only on the univariate case.

The kNCD proposed by \citet{FerratyVieu2003} extends the idea of fkNN such that the estimated probability $\widehat{p}_l, l \in \mathcal{L}$ depends on all training trajectories rather than only on those $k$ closest to ${\bf x}_*$ according to the semi-metric. The computation of $\widehat{p}_l$ in kNCD depends on a kernel function \citep{LiRacine2007}, which weights the training observations by the distance to the new observation given a semi-metric, with larger weights for smaller distances. The new observation is classified into the group with the highest $\widehat{p}_l, l \in \mathcal{L}$. Formally, giving the training observations ${\bf x}_i, \cdots, {\bf x}_n$, and the new observation ${\bf x}_*$, the estimated probability of ${\bf x}_*$ belonging to class $l \in \mathcal{L}$ is defined as:
\begin{eqnarray}\label{eq:posterior2}
\widehat{p}_{l}(h)=\frac{\sum_{i=1}^n {\mathds 1}_{y_i=l}K\left(D({\bf x}_i,{\bf x}_*)/h\right)}{\sum_{i=1}^n K\left(D({\bf x}_i,{\bf x}_*)/h\right)},
\end{eqnarray}
where $K$ is the kernel function and $h > 0$ is the bandwidth parameter that scales the kernel function to the observed data. Note that in expression (\ref{eq:posterior2}), we have again stressed the dependence of the probabilities on the parameter $h$, which needs to be tuned (e.g., cross-validation-based methods), but that we suppress this dependence in the notation for simplicity. Note also that if a Uniform kernel function is considered, expression (\ref{eq:posterior}) is very similar to expression (\ref{eq:posterior2}). In this case, differences in the kNCD method \citep{FerratyVieu2003} and the fkNN \citep{Fuchs2015} only concern the chosen neighbourhood, which has a fixed size according to distance in the first and a fixed number of neighbours in the second case. More details can be found in \citet{FerratyVieu2003, FerratyVieu2006}. 

\subsection{Ensemble} \label{sec:ensemble}

We use ensembles as flexible tools that combine weak learners (e.g., using fkNN or kNCD) based on different semi-metrics chosen to  capture a diverse set of  characteristics of the trajectories. 
Ensemble models are expected to reduce error rates and improve the bias-variance trade-off compared to single weak learners \citep 
{Hastie2009}. We combine weak learners using the stacked generalisation method \citep {Wolpert1992, Breiman1995, LeblancTibshirani1996}, a two-step procedure that uses the predictions of several weak learners (first step) as predictors in a new learning model  (``super-learner", second step), for a categorical output in our classification case. In particular, in addition to stacked generalisation using linear combinations of weak learners' predictions as in \cite{Fuchs2015} \citep[see, for example, the discussion by][]{LeblancTibshirani1996}, we use more flexible super-learners such as tree-based random forests and gradient boosting.

\subsubsection{Stacked generalisation ensembles for functional predictors}\label{sec:ensemble:method}

The first considered ensemble method using linear combinations of weak learner predictions was first used by \citet{Fuchs2015} for univariate functional data classification based on fkNN only. Following \citet{Maierhofer2017}, we call this ensemble LCE (Linear Combination Ensemble), and extend it to the multivariate case with fkNN and kNCD.  

Consider $M$ independent weak learners built on either fkNN or kNCD, each based on a different semi-metric. The first step for LCE building obtains the estimated probabilities $\widehat{p}_{lm}$ 
of the new observation ${\bf x}_*$ belonging to class $l \in \mathcal{L}$ for the $m$-th weak learner for all $m$; see expressions (\ref{eq:posterior}) and (\ref{eq:posterior2}). Note that the $\widehat{p}_{lm}, \forall l,m$ implicitly depend on either the parameter $k$ (if fkNN is used) or $h$ (if kNCD is used) and that both parameters can be tuned with cross-validation methods; see Section \ref{sec:ensemble:practice}.  The second step finds an optimal convex combination of these probabilities. In particular, the probability  $\widehat{\alpha}_{l}$ of the new observation ${\bf x}_*$ belonging to class $l \in \mathcal{L}$ is estimated as: 
\begin{eqnarray}\label{probs:LCE}
\widehat{\alpha}_l=\sum_{m=1}^{M} \omega_m\widehat{p}_{lm},
\end{eqnarray} where the unknown coefficients $\omega_m$ to be estimated satisfy: 
\begin{eqnarray}\label{probs:LCE:restriction}
\omega_m \geq 0 \quad \forall{m}, \quad \sum_{m=1}^{M} \omega_m=1.
\end{eqnarray}

The coefficient $\omega_{m}$ is the contribution (weight or importance) of the $m$-th weak learner (or semi-metric) to the estimated probabilities $\widehat{\alpha}_l, \, \forall l$. The constraint $\omega_m \geq 0, \, \forall m$ ensures that the estimated probabilities $\widehat{\alpha}_l, \, \forall l$ are always positive. Also, the  constraint $\sum_{m=1}^{M}\omega_m=1$  ensures all estimated probabilities $\widehat{\alpha}_l, \, \forall l$ sum up to $1$
\citep[for more details see][]{GertheissTutz2009}. Note that in expression (\ref{probs:LCE}) each weak learner has the same weight for each class as this is the only solution that ensures probability constraints (\ref{probs:LCE:restriction}) are also satisfied for the $\widehat{\alpha}_l$ \citep[see the proof of Proposition 1 by][]{GertheissTutz2009}. 

Coefficients $\omega_m$, $m=1,\cdots, M$ are estimated following \citet{Fuchs2015} on the training sample 
by minimising the Brier score  \citep[][]{Brier1950} subject to  constraints in~(\ref{probs:LCE:restriction}):
\begin{eqnarray}\label{eq:brierscoreG}
S({\boldsymbol \omega})=\frac{1}{n}\sum_{i=1}^n \sum_{l=1}^L \left({\mathds 1}_{y_i=l} -\widehat{\alpha}_{il}\right)^2=\frac{1}{n}\sum_{i=1}^n \sum_{l=1}^{L} \left({\mathds 1}_{y_i=l}-\sum_{m=1}^{M} \omega_m \widehat{p}_{ilm}\right)^2,
\end{eqnarray}
where $\widehat{\alpha}_{il}$ is the estimated probability in  (\ref{probs:LCE}) for the $i$-th  observation. The Brier score is a strictly proper scoring rule, and  the optimiser is thus unique \citep{GneitingRaftery2007}. The constraints in expression (\ref{probs:LCE:restriction}) impose a positive lasso-type penalty \citep{Tibshirani1996} on the coefficients, hence the LCE has the advantage of allowing the selection of weak learners by setting coefficients to zero \citep{Fuchs2015}. 

However, the LCE is restricted to convex linear combinations and equal weights of  weak learners across the different classes. We thus propose to extend the LCE to more flexible combinations of weak learners' predictions using stacked generalisation with more flexible super-learners. Many methods would be possible as super-learners. We focus here on tree-based random forests and tree-based gradient boosting. For univariate functional data classification, the former have been considered before in the Master's thesis of \cite{Maierhofer2017}. Depending on the used super-learner, we will call these ensembles either RFE (Random Forest Ensemble) 
or GBE (Gradient Boosting Ensemble).

Tree-based random forests combine a number of decision trees --weak learners-- tuned on bootstrapped training samples. To reduce between-tree correlation, only a subset of the complete set of predictors is considered in each split in each  tree, ensuring that weaker predictors are also present in some of the trees. As random forest predictions are based on a collection of trees with (ideally) low correlation, such predictions are more reliable --less variable-- than predictions of a single tree. The number of combined trees and the number of predictors to be considered in each tree split are two parameters that need to be tuned using, e.g., cross-validation methods. Tree-based gradient boosting models also combine trees, which in this case are constructed sequentially, with the output of the next tree being the information (pseudo-residuals) that was not explained in the previously grown tree. The number of trees, the number of splits in each tree, and a shrinkage parameter that controls how the gradient boosting model learns are parameters that need to be tuned. 
In classification tasks, both random forest and gradient boosting models assign a new observation to the majority class predicted across all tuned trees. We use here both 
models as super-learners with the estimated probabilities from the selected weak learners as inputs. 
See \citet{Hastie2009,James2013} for more details on random forests and gradient boosting, and Sections \ref{sec:ensemble:practice} and \ref{section:application} for super-learners tuning details. 

RFE and GBE are more flexible methods than LCE: First, they are non-linear classifiers that allow for more complex relationships between the output and the inputs than the linear ones in LCE. Second, random forest and gradient boosting are especially well-suited to dealing with complex interactions between predictors. Because our predictors in this paper are the estimated probabilities given by a collection of weak learners (or semi-metrics), RFE and GBE are better than LCE at capturing potential interactions between semi-metrics. Third, unlike for LCE, RFE and GBE allow different semi-metrics to be important for predicting different classes, which can be important in practice. Another key aspect of RFE and GBE is that they allow including scalar covariates as predictors in addition to the weak learners' estimated probabilities and can even capture interactions between semi-metrics and these covariates. For interpretability, we can use variable importance techniques (e.g., using permutational methods) to measure the overall relevance of each semi-metric for prediction, or class-specific variable importance to identify the most relevant semi-metrics for predicting each of the classes. 

\subsubsection{Stacked generalisation ensembles in practice}\label{sec:ensemble:practice}

Weak learners rely on parameters $k$ or $h$ (cf. Section \ref{sec:weakLearners}). We tune them using nested cross-validation \citep{Stone1974}, which splits the sample into training and validation (inner loop) and testing (outer loop) sub-samples to base model selection (inner loop) and model performance (outer loop) on independent samples, controlling model over-fitting and giving a more honest model performance assessment. 

We use the same sample splits in both the inner and outer loops for all weak learners and all ensembles to ensure fair comparison, with the proportion of correctly classified observations (accuracy) used to measure performance. We first tune the $M$ considered weak learners based either on fkNN or kNCD and a given semi-metric in the inner loop, and evaluate the predictive performance of the tuned learners in the outer loop. This gives estimated probabilities $\widehat{p}_{ilm}, \forall i,l,m$ for all learners $m$, classes $l$ and observations $i$.

We then build the ensembles. For the LCE, linear combination coefficients $\omega_m$ for the $\widehat{p}_{ilm}$ are estimated in the inner loop by solving the optimisation problem in expression (\ref{probs:LCE}). Due to the lasso-type penalty, some of the coefficients $\omega_l$ are set to $0$ and drop out for prediction evaluation in the outer loop.

To tune the super-learners random forest for RFE and gradient boosting for GBE, we add a forward selection step of weak learners to improve predictive performance. First, the two best weak learners are combined in an ensemble with the super-learner tuned on the same inner loop sample splits than for weak learners and LCE. Second, the third best weak learner is included in the ensemble, and the super-learner is re-tuned. It is kept in the ensemble if it improves the inner loop accuracy, then the next best weak learner is considered for inclusion and the process iterated recursively. The final ensemble is then evaluated in terms of outer loop accuracy. 

\section{Application}\label{section:application}

\subsection{Survey design and sample description}

The data used in this application were collected from an online survey conducted by the Institute for Employment Research (IAB) between September and October 2016 \citep{Horwitz2019, FernandezFontelo2021}. $1627$ participants from a previous survey wave in 2014 who had consented to future contact were sent an invitation; of these, $1527$ received a $5$ Euros incentive, while the remaining $100$ were contacted for a pretest of the mouse-tracking software shortly before the beginning of the survey. Finally, $1250$ opened the survey and, of these, $1213$ completed the questionnaire. Section 4.1 of \citet{Horwitz2019} provides more details on the participants' recruiting process.

The survey included four sections with a total of $36$ questions of a variety of formats (see \citeauthor{Horwitz2019} \citeyear{Horwitz2019} for more details); we here focus on a subset of questions for which difficulty was manipulated experimentally, namely ``employment detail", ``employee level" and ``education level" (target questions) from sections ``employment" and ``demography". Of the $1213$ participants who responded and completed the survey, $886$ ($73\%$) used a computer mouse, and $853$ ($70\%$) had continuous mouse-tracking data available. Thus, our sample consists of these $853$ participants, or a subset depending on the survey question. The sample is gender-balanced (51-49\% female-male), with an average age of 51 years. When the survey took place, most of the individuals were employed. 

The survey was only available via the web, and participants were told to answer it using either laptop or desktop computers, as mouse-tracking data could not be collected otherwise. In addition, participants were asked to use a computer mouse if possible. The study was implemented using the SoSci Survey software \citep{Leiner2019}, and we used the mousetrap-web library --a platform-independent, open-source JavaScript library created by \citet{Henninger2016}-- as a tool to collect mouse-tracking data, including the mouse cursor $x$- and $y$-coordinates, timestamps and a representation of the mouse movement trajectory over the computer screen to link the location of the mouse cursor at each time to the question content. We collected mouse movement trajectories from survey respondents who agreed to it at the beginning of the questionnaire for each survey question. Thus raw trajectories between participants within the same question were not necessarily of the same length. 

\subsection{Target questions}

We analysed three different target questions: The first is ``employment detail" (i.e., type of employment), which was manipulated, producing two versions with response options using simple or complex language. The second and third were two ordinal-type questions on the maximum degree of responsibility at work (``employee level") and the degree of education achieved (``education level"), which were manipulated, producing one version with a natural ascending response options' ordering, and another with the options randomly ordered. One version of each target question was randomly assigned to each participant, ensuring that roughly half of the sample received each version of the question. See the Supporting Material for details on each question's layout. 

Employment detail had nine response options, and the majority of the sample was employed in the civil service at the municipality level. Employee level and education detail had respectively four and eleven response options, and most of the respondents carried out a qualified occupation, either following instructions (e.g., accountant) or with some independent activities and responsibilities (e.g., scientific employee), and roughly half of the respondents obtained qualification for university entrance. See the Supporting Material for more information on the response distribution for all three target questions.

\subsection{Mouse movement trajectories}

We applied several pre-processing steps to the mouse movement trajectories. First, coordinates $x$ and $y$ at each timestamp depended on each participant's computer browser; therefore, we standardised them such that each bivariate coordinate pair was comparable across trajectories within the same target question. Second, we computed the first and second derivatives over our raw trajectories separately for each dimension to allow usage of curves ($a=0$), their velocity ($a=1$), and their acceleration ($a=2$). The mouse movement trajectories and their first and second derivatives were then time-normalised to the $[0,1]$ interval. In particular, we used linear interpolation based on the timestamps separately for each trajectory dimension; we used $101$ equidistant time steps for the time-normalisation using the \texttt{R} package \texttt{mousetrap} \citep{mousetrap}. We then applied additional filter criteria detailed in \citet{FernandezFontelo2021}: For example, we removed trajectories from participants who took an exceptionally long time to respond (more than $7$ minutes). We also excluded trajectories from respondents who did not provide an answer for gender or who selected ``other" for gender since only a few participants selected this category or from participants who selected a free-form text input answer option for education level. After these pre-processing steps, $551$, $501$, and $548$ bivariate mouse movement trajectories remained for employment detail (simple vs. complex language), and employee level and education level (sorted vs. unsorted response options), respectively, each of length $101$. These were used as functional predictors of difficulty for each target question. See the Supporting Material for some images of the raw trajectories for each of the three target questions.

Before time normalisation, ten different mouse movement measures grouped in five categories according to their typology were extracted from the raw trajectories with the \texttt{R} package \texttt{mousetrap}: (i) time-type measures such as time from page load until response submission (response time, RT) and time from page load until the first mouse movement occurred (initiation time); (ii) distance-type measures such as the total Euclidean distance travelled by the computer mouse (total distance) (iii) derivative-type measures such as maximum velocity and maximum acceleration; (iv) hover-type measures such as the number of periods without movement exceeding a minimum duration threshold (hovers) and the total time of all periods without movement exceeding a minimum threshold duration (hovers time), and (v) flip-type measures such as the number of changes in movement direction along the horizontal axis (x-flips) and the vertical axis (y-flips). These measures summarising the mouse movement trajectories were previously used for difficulty prediction in \citet{FernandezFontelo2021}. As they also showed that personalisation of measures by correcting for individual baseline behaviours and location of the given answer improved classification, we use personalised measures as defined there. We also considered in our application the total length of the raw mouse movement trajectories (i.e., the total number of timestamps), which is related to the non-personalised response time as the longer the response time, the longer the trajectory.

\subsection{Ensemble models for mouse movement trajectory classification} \label{application:ensembleres}

We considered several ensemble model candidates. First, to combine weak learner predictions, we considered both LCE \citep{Fuchs2015} and the more flexible proposals RFE and GBE (cf. Section \ref{sec:ensemble}). Second, we considered two approaches for including personalised mouse movement measures \citep[see][]{FernandezFontelo2021} into the ensemble models: (i) measure-based semi-metrics (ii) using measures as additional vector-valued predictors in the corresponding super-learner. As this last alternative is only possible for RFE and GBE, this yields five combinations each for both types of weak learners built on fkNN or kNCD --with Gaussian kernel. These ten combinations were tuned and evaluated using nested cross-validation, as described above, using the same $10$ outer (testing sample) and $5$ inner (training and validation samples) folds for all fkNN, kNCD, and ensemble models to ensure a fair comparison. 

The average accuracy across the $10$ testing samples (outer loop) indicated the weak learner's predictive performance. We then selected weak learners with an outer accuracy equal to or greater than $0.55$ for potential inclusion in the ensemble. We selected $24$ and $18$ weak learners for fkNN and kNCD, respectively, for employment detail, $12$ and $12$ for employee level, and $6$ and $10$ for education level. Between $2$ and $6$ weak learners were built on measure-based semi-metrics (if ensembles are built based on approach (i)). 

We then constructed ensembles based on either LCE, RFE or GBE. For LCE, following \citet{Fuchs2015}, we estimated the linear combination coefficients $\omega_m$, $m=1, \cdots, M$, using the function \texttt{lsei} from the \texttt{R} package \texttt{limSolve} to solve the optimisation problem with constraint (\ref{probs:LCE:restriction}), resulting in a lasso-type penalty (cf. Section \ref{sec:ensemble:practice}). Note that the probabilities ${\hat p}_{ilm}$ for the different weak learners in (\ref{eq:brierscoreG}) were computed on the same inner splits used to estimate the LCE coefficients, and the LCE's accuracy was then evaluated on the testing sample. For RFE and GBE construction, we had to tune super-learners. For the random forest in RFE, we used the estimated probabilities ${\hat p}_{ilm}$ for observations in the inner loop sample as random forest predictors, and tuned the random forest parameters (number of trees and predictors in each tree) using $10$-fold cross-validation and maximising the inner loop's accuracy. RFE's predictive performance was evaluated with the average accuracy over the $10$ testing samples (outer loop). Note that for nested cross-validation, super-learners are differently tuned for each sample split in the outer loop. We used an analogous procedure for tuning (number of trees, shrinkage parameter, and a parameter related to the level of interaction between predictors) and performance evaluation of GBE.

Table \ref{app:tab2} provides the outer loop accuracies (those greater than $0.55$) of the weak learners built on kNCD for two of our target questions (employment detail: simple vs. complex language, and employee level: ordered vs. unordered response options). For employment detail, chosen semi-metrics included most notably those in the scalar-vector-summary (svs) family, such as different types of maximums and ranges as well as a mean-type semi-metric (some based on the first ($a=1$) or the second ($a=2$) derivatives), and semi-metrics based on personalised measures such as the Euclidean distance between response times (RT), hovers and hover time and vertical flips ($y$-flips), as well as the Euclidean distance between the bivariate vectors of flips (x- and y-flips) (bivariate flips). We also included the Hausdorff distance (elastic family), and semi-metrics from the lock-step family such as different versions of $L^p$ distances and a correlation-type distance (based on the second derivative of the raw trajectory, $a=2$).
The elastic family was more important for employee level than employment detail since DTW, Fr\'echet and Hausdorff distances were relevant for difficulty prediction in this second target question. Also, semi-metrics for employee level included different types of minimums and ranges, semi-metrics based on personalised measures such as the Euclidean distance between initiation times and the number of hovers between trajectories (svs family), and a correlation-type distance (lock-step family). Finally, for education level, we selected semi-metrics mainly from the svs family, including distances between means, maximums and minimums in both $x$ and $y$-direction. We also considered different versions of the $L^p$ distance from the lock-step family, DTW (elastic family) and Hamming distance (symbol sequence type family). If weak learners were instead based on fkNN, we selected similar semi-metrics than for kNCD for each target question. Although other elastic-type distances have been considered in Section \ref{methods:semi} \citep[e.g.,][]{Steyer2021, Srivastava2016}, these have not been explored in this application due to their computing cost. 

Table \ref{app:tab3} shows the inner loop accuracies for weak learner selection in kNCD ensemble construction and the outer loop accuracies for evaluating the predictive capacities of the selected kNCD ensembles for employment detail and employee level questions. Results for fkNN ensembles were similar (although slightly less good and/or consistent) and are given in the Supporting Material together with similar results for education level.

The best kNCD ensemble model for employment detail used gradient boosting as a super-learner and included mouse movement measures in the ensemble via measure-based semi-metrics. Ensembles added new weak learners in a forward step-wise fashion if they improved the inner loop accuracy (cf. Section \ref{sec:ensemble:practice}). Accuracies in bold indicate the inclusion of the corresponding weak learner; the ensemble model for employment detail contained measure-based semi-metrics for scalar measures such as response times (RT), trajectory lengths (length), number of hovers, time hovering and vertical flips, and the $\mathbb{R}^2$ Euclidean distance between vectors of horizontal and vertical flips (bivariate flips). The ensemble model also included other semi-metrics from the svs family, such as the maximum distance in the $y$ (globMax-y) and $x$ (globMax-x) coordinates separately computed on the time-normalised trajectories ($a=0$), the mean distance computed on the time-normalised first derivative of the raw trajectories ($a=1$, velocity), and from the lock-step family such as the correlation-type distance computed on the time-normalised trajectories. We selected this ensemble model with an inner loop accuracy of $0.7057$ and an outer loop accuracy --in italics and bold-- of $0.6825$.

Keeping the splits in the inner and outer loops constant across methods not only allows a fair comparison of accuracies in Tables \ref{app:tab2} and \ref{app:tab3}, but also with our previous work \citep{FernandezFontelo2021}, where we analysed the predictive performance of a set of classifiers based only on scalar mouse movement measures as predictors. In particular, \citet{FernandezFontelo2021} found that the best model for employment detail was a tree-based gradient boosting with personalised measures, obtaining an outer loop accuracy of $0.6587$. The three most relevant mouse movement measures in that model were response time, y-flips and x-flips. While these were also selected for potential inclusion in the ensemble (Table \ref{app:tab2}), the outer loop accuracy $0.6825$ showed better predictive performance of the ensemble model than \citet{FernandezFontelo2021}.

The best kNCD ensemble model for employee level was a stacked-generalisation-type ensemble model with a gradient boosting model as a super-learner. This ensemble model was smaller, combining only three weak learners with semi-metrics from the elastic (DTW) and svs (globMin-y and globRange-y) families and personalised mouse movement measures as scalar-vector predictors in the super-learner. We selected this ensemble model with an inner loop accuracy of $0.7788$ and an outer loop accuracy of $0.7664$. In comparison, \citet{FernandezFontelo2021} found that the best model for employee level was gradient boosting with personalised measures, obtaining an outer loop accuracy of only $0.5909$. In that model, the three most important measures were initiation time, y-flips, and the number of hovers, and two of these were also relevant here. However, the outer loop accuracy of $0.7664$ for the ensemble model significantly improves the outer-loop accuracy of $0.5909$ of \citet{FernandezFontelo2021}. For this second target question, the distance measure that worked best for prediction was DTW, commonly used to compare trajectories with different speeds. 

In general, it appears that semi-metrics from the svs family worked best for predicting difficulty in each of the two target questions. In particular, semi-metrics such as differences in maximum values in either the first or second dimension separately and several measure-based semi-metrics worked particularly well for predicting difficulty in employment detail (easy vs. complex language). Another semi-metric that performed quite well for prediction in this target question was the difference in the trajectories' mean velocities. Unlike for employment detail, in employee level, the best semi-metric for difficulty prediction (ordered vs. unordered answering options) was DTW, an elastic-type distance. In terms of prediction, the distance between minimums and ranges in the second dimension was likewise important for this target question. Differences in question manipulations and resulting types of question difficulty can explain the differences in best-performing semi-metrics between both target questions. For employment detail, for example, longer response times and lower velocities are expected in the difficult setting with complex language. Trajectories for employee level, on the other hand, are expected to have more accelerations and decelerations for the unordered setting, given that participants with difficulties are expected to go up and down with the mouse to find the correct answer more frequently, and this is well caught by distances such as DTW.

\begin{felixhack}
\begin{table}
\caption{\label{app:tab2} Accuracies measuring predictive performance (outer loop) of our selection of weak learners built on kNCD for two of the three target questions (employment detail and employee level). Weak learners were chosen if outer-loop accuracy was equal to or greater than $0.55$, and in this case used as candidates for ensemble construction (see Table \ref{app:tab3}).}
\centering
\fbox{%
\begin{tabular}{lllclllc}
\toprule
\multicolumn{4}{c}{\textbf{Employment detail}}&
\multicolumn{4}{c}{\textbf{Employee level}}\\
\cmidrule(r){1-4}\cmidrule(r){5-8}
{\bf Family} & {\bf Semi-metric } & $\boldsymbol{a}$ & {\bf accuracy }  & {\bf Family } & {\bf Semi-metric } & $\boldsymbol{a}$ & {\bf accuracy } \\
\midrule
svs & RT &  & .6135 & elastic & DTW & 0 & .7286 \\
 & globMax-y & 0 & .6044  & svs & globMin-y & 0 & .7085 \\
 & length &  &  .6008 &  & globMin & 0 & .7005 \\
 & globMax-x & 0 & .5952  &  & globRange-y & 0 & .6029 \\
 & hovers &  & .5863  &  & globRange & 0 & .5927 \\
& y-flips &  & .5843  & & initiation time & & .5830 \\
elastic & Hausdorff & 0 & .5753  & lock-step & $L^1$ & 0 & .5807 \\
svs & mean & 1 &  .5699 & elastic & Fr\'echet & 0 & .5749 \\
 & hover time &  & .5699 & lock-step & correlation & 1 & .5707 \\
 & globMax & 0 & .5697  & elastic & Hausdorff & $0$ & .5651 \\
lock-step & correlation & 0 & .5681  & svs & globMax & $2$ & .5629 \\
svs & globMax-y & 1 & .5572  &  & hovers & & .5528  \\
& globRange-y & 1 & .5572  &  &  & &  \\
lock-step & $L^4$ & 0 & .5555  &  &  & &  \\
& $L^1$ & 0 & .5554  &  &  & &  \\
svs & globMax-y & 2 & .5535  &  &  & &  \\
 & bivariate flips &  & .5534 &  &  & &  \\
 & globMax & 2 & .5518  &  &  & &  \\
\hline
\end{tabular}}
\end{table}
\end{felixhack}

Outer loop accuracies show better performance of the more flexible ensembles RFE and GBE than the LCE. For employment detail, the outer loop accuracies of the four ensembles with tree-based super-learners in Table \ref{app:tab3} are much higher than those of the LC ensemble, which is even worse than that obtained in \citet{FernandezFontelo2021}. For employee level, the advantage of the tree-based super-learners over the LCE is similar but not as pronounced. Overall, it seems that gradient boosting worked better than random forest as a super-learner as it was selected for each of the three target questions in kNCD and for education level in fkNN; see Table \ref{app:tab3} and the Supporting Material. 

The Supporting Material gives examples of the weak learners' outer loop accuracies for the three target questions for fkNN and kNCD. The Supporting Material also contains inner loop accuracies for weak learners (fkNN or kNCD) selection in RF (type I and II) and GB (type I and II) ensembles and the outer loop accuracies for the LCE, RFE, and GBE. Finally, the Supporting Material also provides details and the link to a GitHub repository, which contains further information on the inner and outer loops accuracies for the weak learners, LCE, RFE, and GBE, and the estimated LCE coefficients for the three target questions.

\begin{table}
\caption{\label{app:tab3} Inner-loop accuracies for kNCD ensembles for which one weak learner (row-wise) was added at a time and included if it improved the accuracy (in bold). Type I ensembles included personalised measures following \citet{FernandezFontelo2021} in the ensemble using measure-based semi-metrics, and type II included these measures as scalar-vector covariates in the super-learner. Used super-learners were either a random forest (RF), gradient boosting (GB), or linear combinations (LC). Accuracies in italics indicate outer loop predictive performance of the total ensemble, including all semi-metrics with bolded rows. Accuracies in both bold and italics are the outer loop accuracies for the selected ensembles for each target question.}
\centering
\fbox{%
\begin{tabular}{lllcccccc}
\toprule
&&&&\multicolumn{2}{c}{\textbf{type I}}& \multicolumn{2}{c}{\textbf{type II}} &\\
\cmidrule(r){5-6}\cmidrule(r){7-8}
{\bf Target question} & {\bf Family} & {\bf Semi-metric} & $\boldsymbol{a}$ & {\bf RF}  & {\bf GB} & {\bf RF} & {\bf GB} & {\bf LC} \\
\hline 
 & svs & RT & & & & - & - & \\
{\bf Employment detail} &  & globMax-y & 0 & {\bf .6437} & {\bf .6472} & & &\\
(simple/complex &  & length & & .6427 & {\bf .6548} & - & - &\\
language) &  & globMax-x & 0 & {\bf .6657} & {\bf .6879} & {\bf .6820} &  {\bf .6924} & \\
&  & hovers & & {\bf .6708} & {\bf .6885} & - & - & \\
& & y-flips & & {\bf .6765} & {\bf .6937}& - & - & \\
& elastic & Hausdorff & 0 & .6755 &.6901& .6786 & .6877 & \\
& svs & mean & 1 & .6741 & {\bf .6946}& {\bf .6824}& .6923 & \\
&  & hover time & & {\bf .6793} & {\bf .7020}& - & - & \\
&  & globMax & 0 & .6751 & .6986 & .6832 & .6918 & \\
& lock-step & correlation & 0 & {\bf .6799} & {\bf .7049} & {\bf .6884}& {\bf .6978} & \\
& svs & globMax-y & 1 & {\bf .6801} & .7038 & {\bf .6896}& {\bf .7033} & \\
& & globRange-y & 1 & .6794 & .7032 & .6880 & .7030 & \\
& lock-step & $L^4$ & 0 & .6768 & .7033 & .6853 & .7002 & \\
& & $L^1$ & 0 & {\bf .6828}& .7010 & .6871 & .7029 & \\
& svs & globMax-y &2& {\bf .6829}& .6982 & {\bf .6900} & .6986 & \\
&  & bivariate flips&  & .6793 & {\bf .7057} & - & - & \\
&  & globMax &2& .6812 & .6980 & .6884 & .6984 & \\
\cmidrule(r){2-9}
& & & & {\it .6806} & \textit{\textbf{.6825}}& {\it .6878} & {\it .6951} & {\it .5972} \\
\hline
& elastic & DTW & 0 &&&&& \\
{\bf Employee level} & svs &  globMin-y & 0 & {\bf .7567} & {\bf .7664}& {\bf .7767} & {\bf .7773}&\\
(ordered/unordered  &  & globMin & 0 & {\bf .7576} & .7611 & .7762& .7737 & \\
 response options)& & globRange-y & 0 & {\bf .7667} & .7664 & .7743 & {\bf .7788} & \\
& & globRange & 0 & .7625 & .7656 & .7725  & .7764 & \\
& & initiation time & & {\bf .7671}& {\bf .7737} & - & - & \\
& lock-step & $L^1$ & 0 & .7658 & .7727& .7753 & .7753 & \\
& elastic & Fr\'echet & 0 & .7627 & .7727 & .7738  & .7782 &\\
& lock-step & correlation & 1 & .7654 & .7720 & .7753 & .7755 & \\
& elastic & Hausdorff & 0 & .7653 & .7687 & .7729 & .7755 & \\
& svs & globMax & 2 & .7667 &  .7720 &.7696 & .7786 & \\
&  & hovers & & {\bf .7718}&{\bf .7758} &-&-&\\
\cmidrule(r){2-9}
& & & & {\it .7605} & {\it .7784} & {\it .7924} & \textbf{\textit{.7664}}& {\it .7205} \\
\end{tabular}}
\end{table}

\section{Discussion} \label{section:discussion}

Computer mouse movements as a source of paradata have been used in survey research to improve different survey aspects, including survey data quality \citep{Kreuter2013, Horwitz2017}. For web surveys, it has been shown that a number of features of these mouse movements (mouse movement measures) were statistically related to question difficulty \citep{Horwitz2017, Horwitz2019} and that several measures were good predictors of question difficulty \citep{FernandezFontelo2021}. In terms of web survey data quality, the detection of difficulty in, e.g., specific survey questions or participants, may help identify potential sources of measurement errors in responses as well as data quality issues \citep{Yan2013}. However, to date, the potential of using the entire mouse movement trajectory as bivariate functional predictors has not yet been fully explored. To this end, we analysed here data from a web survey that included a number of experimentally manipulated questions to create two levels of difficulty, and collected mouse movement trajectories for each of these questions (target questions). We thus investigated the potential of the trajectories as functional predictors to identify when respondents faced difficulties with specific questions, showing that mouse movement curves as a whole contain more information for prediction than mouse movement measures alone.

Many existing methods that use functional predictors can be classified into approaches that reduce the dimension by extracting relevant features of the functions and using these in machine learning models \citep [e.g.,][]{FernandezFontelo2021}, and into genuinely functional approaches using whole functions as predictors. The main limitation of the first approach is that it summarises the functional observations, which may result in a loss of valuable information. Methods in the second class address this point, but are most often designed for univariate functional predictors. The small fraction of these methods that also allow for multivariate functional predictors rarely use ensemble methods to improve models' predictive performance, and if they do combine weak learners only based on a single semi-metric such as $L^2$ or DTW, thus capturing only certain kinds of information contained in these functions.

Due to the limitations of existing classification methods for multivariate functional predictors, and in light of our applications' goal of predicting question difficulty in web surveys using mouse movement trajectories (bivariate functions), we introduced here new ensemble models for multivariate functional predictors as extensions of the non-parametric models fkNN \citep{Fuchs2015} and kNCD \citep{FerratyVieu2003, FerratyVieu2006}. In particular, our ensemble models combine a broad collection of weak learners built on fkNN or kNCD, which are in turn based on different semi-metrics. These weak learners are combined using stacked generalisation techniques, using as super-learners either random forest (RFE) or gradient boosting (GBE). 

RFE and GBE improve on existing methods as follows: First, they are genuinely functional methods that allow for both univariate and multivariate functional predictors. For example, when compared to \citet{FernandezFontelo2021}, which used only features of the trajectories as predictors in a set of typical machine learning methods, our models show significantly better results as they consider information in the entire bivariate function. Second, RFE and GBE are built on a broad collection of semi-metrics, each of which captures a different aspect of our functional observations; thus, they are more comprehensive than methods considering a single semi-metric \citep[e.g.,][]{Selk2021,Prieto2013,Grecki2015,Mei2016,Wang2017}. In fact, our results (Tables \ref{app:tab2} and \ref{app:tab3}) show that neither $L^2$ nor DTW are always the most important semi-metric in our application. Third, they combine different sources of information using stacked generalisation methods, which can improve the predictive performance compared to non-ensemble-based models, but have rarely been used in the functional classification setting so far \citep[e.g.,][]{Prieto2013, Wang2017}.
 
In our application, we analysed three target questions in a web survey conducted by the IAB and found that our functional approach significantly improved predictions on all questions, with up to 18 percentage point improvement (59\% to 77\%) in predictive accuracy (on the employee level item) compared to machine learning methods using only scalar summary measures. Gradient boosting based on trees was the best super-learner in our application. The most important semi-metrics were based on scalar summaries and elastic-type distances, but other semi-metrics also added information. We also saw differences in the best-performing semi-metrics depending on the difficulty manipulation for a given question, likely related to the different reactions caused by these difficulties.  

Although our models performed well in prediction, some potential improvements remain yet unexplored. For instance, while we include elastic semi-metrics that allow stretching and compressing time, these still focus on the whole function and keep the temporal ordering. Therefore, it may also be useful in future research to develop methods that look at more local characteristics, subsections, and interactions of trajectories, or to include other weak learners than fkNN and kNCD. 

\bibliographystyle{rss}
\bibliography{Fernandez-Fontelo}

\begin{thebibliography}{62}
\expandafter\ifx\csname natexlab\endcsname\relax\def\natexlab#1{#1}\fi
\expandafter\ifx\csname url\endcsname\relax
  \def\url#1{\texttt{#1}}\fi
\expandafter\ifx\csname urlprefix\endcsname\relax\def\urlprefix{URL: }\fi

\bibitem[{Abanda et~al.(2019)Abanda, Mori and Lozano}]{Abanda2018}
Abanda, A., Mori, U. and Lozano, J.~A. (2019) A review on distance based time
  series classification.
\newblock \textit{Data Mining and Knowledge Discovery}, \textbf{33}, 378--412.

\bibitem[{Alt and Guibas(2000)}]{Alt2000}
Alt, H. and Guibas, L.~J. (2000) Chapter 3 - discrete geometric shapes:
  Matching, interpolation, and approximation.
\newblock In \textit{Handbook of Computational Geometry}, 121--153.
  North-Holland.

\bibitem[{Bagnall et~al.(2016)Bagnall, Lines, Bostrom, Large and
  Keogh}]{Bagnall2016}
Bagnall, A., Lines, J., Bostrom, A.~G., Large, J. and Keogh, E.~J. (2016) The
  great time series classification bake off: a review and experimental
  evaluation of recent algorithmic advances.
\newblock \textit{Data Mining and Knowledge Discovery}, \textbf{31}, 606--660.

\bibitem[{Breiman(1996)}]{Breiman1995}
Breiman, L. (1996) Stacked regressions.
\newblock \textit{Machine Learning}, \textbf{24}, 49--64.

\bibitem[{Brier(1950)}]{Brier1950}
Brier, G.~W. (1950) Verification of forecasts expressed in terms of
  probability.
\newblock \textit{Monthly Weather Review}, \textbf{78}, 1--3.

\bibitem[{Brockhaus et~al.(2020)Brockhaus, R\"ugamer and
  Greven}]{Brockhaus2020}
Brockhaus, S., R\"ugamer, D. and Greven, S. (2020) Boosting functional
  regression models with fdboost.
\newblock \textit{Journal of Statistical Software}, \textbf{94}, 1--50.

\bibitem[{Callegaro(2013)}]{Callegaro2013}
Callegaro, M. (2013) \textit{Improving Surveys with Paradata: Paradata in Web
  Surveys}, chap.~11, 259--279.
\newblock John Wiley \& Sons, Ltd.

\bibitem[{Chen et~al.(2001)Chen, Anderson and Sohn}]{Chen2001}
Chen, M.~C., Anderson, J.~R. and Sohn, M.~H. (2001) What can a mouse cursor
  tell us more? {Correlation} of eye/mouse movements on web browsing.
\newblock In \textit{CHI '01 Extended Abstracts on Human Factors in Computing
  Systems}, 281--282.

\bibitem[{Conrad et~al.(2007)Conrad, Schober and Coiner}]{Conrad2007}
Conrad, F.~G., Schober, M.~F. and Coiner, T. (2007) Bringing features of human
  dialogue to web surveys.
\newblock \textit{Applied Cognitive Psychology}, \textbf{21}, 165--187.

\bibitem[{Couper(2000)}]{Couper2000}
Couper, M.~P. (2000) Usability evaluation of computer-assisted survey
  instruments.
\newblock \textit{Social Science Computer Review}, \textbf{18}, 384--396.

\bibitem[{Couper(2017)}]{Couper2017}
--- (2017) Birth and diffusion of the concept of paradata.
\newblock \textit{Advences in Social Research \& Japanese Association for
  Social Research}, \textbf{18}, 14--26.

\bibitem[{Dempster et~al.(2020)Dempster, Petitjean and Webb}]{Dempster2020}
Dempster, A., Petitjean, F. and Webb, G.~I. (2020) {ROCKET}: exceptionally fast
  and accurate time series classification using random convolutional kernels.
\newblock \textit{Data Mining and Knowledge Discovery}, \textbf{34},
  1454--1495.

\bibitem[{Fern\'andez-Fontelo et~al.(2021)Fern\'andez-Fontelo, Kieslich,
  Henninger, Kreuter and Greven}]{FernandezFontelo2021}
Fern\'andez-Fontelo, A., Kieslich, P.~J., Henninger, F., Kreuter, F. and
  Greven, S. (2021) Predicting question difficulty in web surveys: A machine
  learning approach based on mouse movement features.
\newblock \textit{Social Science Computer Review}.

\bibitem[{Ferraty and Vieu(2000)}]{FerratyVieu2000}
Ferraty, F. and Vieu, P. (2000) Dimension fractale et estimation de la
  r\'egression dans des espaces vectoriels semi-norm\'es.
\newblock \textit{Comptes Rendus de l'Acad\'emie des Sciences - Series I -
  Mathematics}, \textbf{330}, 139--142.

\bibitem[{Ferraty and Vieu(2003)}]{FerratyVieu2003}
--- (2003) Curves discrimination: a nonparametric functional approach.
\newblock \textit{Computational Statistics \& Data Analysis}, \textbf{44},
  161--173.

\bibitem[{Ferraty and Vieu(2006)}]{FerratyVieu2006}
--- (2006) \textit{Nonparametric Functional Data Analysis: Theory and
  Practice}.
\newblock Berlin, Heidelberg: Springer-Verlag, 1 edn.

\bibitem[{Filzmoser et~al.(2018)Filzmoser, Hron and Templ}]{Filzmoser2018}
Filzmoser, P., Hron, K. and Templ, M. (2018) \textit{Applied Compositional Data
  Analysis With Worked Examples in R}.
\newblock Berlin, Heidelberg: Springer-Verlag, 1 edn.

\bibitem[{Freeman(2018)}]{Freeman2018}
Freeman, J.~B. (2018) Doing psychological science by hand.
\newblock \textit{Current Directions in Psychological Science}, \textbf{27},
  315--323.

\bibitem[{Fuchs et~al.(2015)Fuchs, Gertheiss and Tutz}]{Fuchs2015}
Fuchs, K., Gertheiss, J. and Tutz, G. (2015) Nearest neighbor ensembles for
  functional data with interpretable feature selection.
\newblock \textit{Chemometrics and Intelligent Laboratory Systems},
  \textbf{146}, 186--197.

\bibitem[{Gertheiss and Tutz(2009)}]{GertheissTutz2009}
Gertheiss, J. and Tutz, G. (2009) Feature selection and weighting by nearest
  neighbor ensembles.
\newblock \textit{Chemometrics and Intelligent Laboratory Systems},
  \textbf{99}, 30--38.

\bibitem[{Gneiting and Raftery(2007)}]{GneitingRaftery2007}
Gneiting, T. and Raftery, A.~E. (2007) Strictly proper scoring rules,
  prediction, and estimation.
\newblock \textit{J. Am. Statist. Ass.}, \textbf{102}, 359--377.

\bibitem[{G{\'o}recki and Luczak(2015)}]{Grecki2015}
G{\'o}recki, T. and Luczak, M. (2015) Multivariate time series classification
  with parametric derivative dynamic time warping.
\newblock \textit{Expert Systems with Applications}, \textbf{42}, 2305--2312.

\bibitem[{Greven and Scheipl(2017)}]{Greven2017}
Greven, S. and Scheipl, F. (2017) A general framework for functional regression
  modelling.
\newblock \textit{Statistical Modelling}, \textbf{17}, 1--35.

\bibitem[{Hastie et~al.(2009)Hastie, Tibshirani and Friedman}]{Hastie2009}
Hastie, T., Tibshirani, R. and Friedman, J.~H. (2009) \textit{The elements of
  statistical learning: data mining, inference, and prediction}.
\newblock New-York: Springer, 2 edn.

\bibitem[{Heerwegh(2003)}]{Heerwegh2003}
Heerwegh, D. (2003) Explaining response latencies and changing answers using
  client-side paradata from a web survey.
\newblock \textit{Social Science Computer Review}, \textbf{21}, 360--373.

\bibitem[{Henninger and Kieslich(2016)}]{Henninger2016}
Henninger, F. and Kieslich, P.~J. (2016) Beyond the lab: Collecting
  mouse-tracking data in online studies.
\newblock In \textit{Abstracts of the 58th Conference of Experimental
  Psychologists}.

\bibitem[{Horwitz et~al.(2020)Horwitz, Brockhaus, Henninger, Kieslich,
  Schierholz, Keusch and Kreuter}]{Horwitz2019}
Horwitz, R., Brockhaus, S., Henninger, F., Kieslich, P.~J., Schierholz, M.,
  Keusch, F. and Kreuter, F. (2020) \textit{Learning from Mouse Movements:
  Improving Questionnaires and Respondents' User Experience Through Passive
  Data Collection}, chap.~16, 403--425.
\newblock John Wiley \& Sons, Ltd.

\bibitem[{Horwitz et~al.(2017)Horwitz, Kreuter and Conrad}]{Horwitz2017}
Horwitz, R., Kreuter, F. and Conrad, F. (2017) Using mouse movements to predict
  web survey response difficulty.
\newblock \textit{Social Science Computer Review}, \textbf{35}, 388--405.

\bibitem[{James et~al.(2013)James, Witten, Hastie and Tibshirani}]{James2013}
James, G., Witten, D., Hastie, T. and Tibshirani, R. (2013) \textit{An
  Introduction to Statistical Learning: with Applications in R}.
\newblock Springer New York, 1 edn.

\bibitem[{James(2002)}]{James2002}
James, G.~M. (2002) Generalized linear models with functional predictors.
\newblock \textit{J. R. Statist. Soc. B}, \textbf{64}, 411--432.

\bibitem[{Kieslich et~al.(2021)Kieslich, Wulff, Henninger, Haslbeck and
  Brockhaus}]{mousetrap}
Kieslich, P.~J., Wulff, D.~U., Henninger, F., Haslbeck, J. M.~B. and Brockhaus,
  S. (2021) \textit{mousetrap: Process and Analyze Mouse-Tracking Data}.
\newblock \urlprefix\url{https://CRAN.R-project.org/package=mousetrap}.

\bibitem[{Kreuter(2013)}]{Kreuter2013}
Kreuter, F. (2013) \textit{Improving Surveys with Paradata: Introduction},
  chap.~1, 1--9.
\newblock John Wiley \& Sons, Ltd.

\bibitem[{Kreuter and Casas-Cordero(2010)}]{Kreuter2010}
Kreuter, F. and Casas-Cordero, C. (2010) Paradata.
\newblock \textit{RatSWD Working Papers 136}, German Data Forum (RatSWD).

\bibitem[{Leblanc and Tibshirani(1996)}]{LeblancTibshirani1996}
Leblanc, M. and Tibshirani, R. (1996) Combining estimates in regression and
  classification.
\newblock \textit{J. Am. Statist. Ass.}, \textbf{91}, 1641--1650.

\bibitem[{Leiner(2019)}]{Leiner2019}
Leiner, D.~J. (2019) \textit{SoSci Survey (Version 3.1.06)}.
\newblock \urlprefix\url{Available online from: https://www.soscisurvey.de.}

\bibitem[{Li and Racine(2007)}]{LiRacine2007}
Li, Q. and Racine, J.~S. (2007) \textit{Nonparametric Econometrics : Theory and
  Practice}.
\newblock Princeton University Press, 1 edn.

\bibitem[{Maierhofer(2017)}]{Maierhofer2017}
Maierhofer, T. (2017) Classification of functional data - interpretable
  ensemble approaches.
\newblock Master Thesis, LMU Munich.

\bibitem[{Marron(2021)}]{Marron2021}
Marron, J.~S. (2021) \textit{Object oriented data analysis}.
\newblock Chapman \& Hall/CRC, 1 edn.

\bibitem[{Marx and Eilers(1999)}]{Marx1999}
Marx, B.~D. and Eilers, P. H.~C. (1999) Generalized linear regression on
  sampled signals and curves: A p-spline approach.
\newblock \textit{Technometrics}, \textbf{41}, 1--13.

\bibitem[{McLean et~al.(2014)McLean, Hooker, Staicu, Scheipl and
  Ruppert}]{McLean2014}
McLean, M.~W., Hooker, G., Staicu, A.-M., Scheipl, F. and Ruppert, D. (2014)
  Functional generalized additive models.
\newblock \textit{Journal of Computational and Graphical Statistics},
  \textbf{23}, 249--269.

\bibitem[{Mei et~al.(2016)Mei, Liu, fang Wang and Gao}]{Mei2016}
Mei, J., Liu, M., fang Wang, Y. and Gao, H. (2016) Learning a mahalanobis
  distance-based dynamic time warping measure for multivariate time series
  classification.
\newblock \textit{IEEE Transactions on Cybernetics}, \textbf{46}, 1363--1374.

\bibitem[{Mittereder and West(2021)}]{Mittereder2021}
Mittereder, F. and West, B.~T. (2021) A dynamic survival modeling approach to
  the prediction of web survey breakoff.
\newblock \textit{Journal of Survey Statistics and Methodology}, 1--34.

\bibitem[{Mori et~al.(2016)Mori, Mendiburu and Lozano}]{Mori2016}
Mori, U., Mendiburu, A. and Lozano, J.~A. (2016) Distance measures for time
  series in r: The tsdist package.
\newblock \textit{The R Journal}, \textbf{8}, 451--459.

\bibitem[{Morris(2015)}]{Morris2015}
Morris, J.~S. (2015) Functional regression.
\newblock \textit{Annual Review of Statistics and Its Application}, \textbf{2},
  321--359.

\bibitem[{M\"uller and Stadtm\"uller(2005)}]{Muller2004}
M\"uller, H.-G. and Stadtm\"uller, U. (2005) {Generalized functional linear
  models}.
\newblock \textit{Ann. Statist.}, \textbf{33}, 774--805.

\bibitem[{Navarro(2001)}]{Navarro99}
Navarro, G. (2001) A guided tour to approximate string matching.
\newblock \textit{ACM Computing Surveys}, \textbf{33}, 31--88.

\bibitem[{O'Hora et~al.(2016)O'Hora, Carey, Kervick, Crowley and
  Dabrowski}]{Ohora2016}
O'Hora, D., Carey, R., Kervick, A., Crowley, D. and Dabrowski, M. (2016)
  Decisions in motion: Decision dynamics during intertemporal choice reflect
  subjective evaluation of delayed rewards.
\newblock \textit{Scientific Reports}, \textbf{6}, 20740.

\bibitem[{Pfisterer et~al.(2019)Pfisterer, Beggel, Sun, Scheipl and
  Bischl}]{Pfisterer2019}
Pfisterer, F., Beggel, L., Sun, X., Scheipl, F. and Bischl, B. (2019)
  Benchmarking time series classification - functional data vs machine learning
  approaches.
\newblock \urlprefix\url{https://arxiv.org/abs/1911.07511}.

\bibitem[{Prieto et~al.(2013)Prieto, Alonso and Diez}]{Prieto2013}
Prieto, O.~J., Alonso, C.~J. and Diez, J. J.~R. (2013) Stacking for
  multivariate time series classification.
\newblock \textit{Pattern Analysis and Applications}, \textbf{18}, 297--312.

\bibitem[{Rizzo and Sz\'ekely(2016)}]{energy}
Rizzo, M. and Sz\'ekely, G. (2016) \textit{energy: E-Statistics: Multivariate
  Inference via the Energy of Data}.
\newblock \urlprefix\url{https://CRAN.R-project.org/package=energy}.

\bibitem[{Selk and Gertheiss(2021)}]{Selk2021}
Selk, L. and Gertheiss, J. (2021) Nonparametric regression and classification
  with functional, categorical, and mixed covariates.
\newblock \urlprefix\url{https://arxiv.org/abs/2111.03115}.

\bibitem[{Srivastava and Klassen(2016)}]{Srivastava2016}
Srivastava, A. and Klassen, E.~P. (2016) \textit{Functional and shape data
  analysis}.
\newblock New York: Springer, 1 edn.

\bibitem[{Steyer et~al.(2021)Steyer, Stocker and Greven}]{Steyer2021}
Steyer, L., Stocker, A. and Greven, S. (2021) Elastic analysis of irregularly
  or sparsely sampled curves.
\newblock \urlprefix\url{https://arxiv.org/abs/2104.11039}.

\bibitem[{Stieger and Reips(2010)}]{Stieger2010}
Stieger, S. and Reips, U.-D. (2010) What are participants doing while filling
  in an online questionnaire: A paradata collection tool and an empirical
  study.
\newblock \textit{Computers in Human Behavior}, \textbf{26}, 1488--1495.

\bibitem[{Stillman et~al.(2017)Stillman, Medvedev and Ferguson}]{Stillman2017}
Stillman, P.~E., Medvedev, D.~S. and Ferguson, M.~J. (2017) Resisting
  temptation: Tracking how self-control conflicts are successfully resolved in
  real time.
\newblock \textit{Psychological Science}, \textbf{28}, 1240--1258.

\bibitem[{Stone(1974)}]{Stone1974}
Stone, M. (1974) Cross-validatory choice and assessment of statistical
  predictions.
\newblock \textit{J. R. Statist. Soc. B}, \textbf{36}, 111--133.

\bibitem[{Sz\'ekely et~al.(2007)Sz\'ekely, Rizzo and Bakirov}]{Gabor2007}
Sz\'ekely, G.~J., Rizzo, M.~L. and Bakirov, N.~K. (2007) Measuring and testing
  dependence by correlation of distances.
\newblock \textit{Ann. Statist.}, \textbf{35}, 2769--2794.

\bibitem[{Tibshirani(1996)}]{Tibshirani1996}
Tibshirani, R. (1996) Regression shrinkage and selection via the lasso.
\newblock \textit{J. R. Statist. Soc. B}, \textbf{58}, 267--288.

\bibitem[{Wang and Wu(2017)}]{Wang2017}
Wang, H. and Wu, J. (2017) Boosting algorithm for real-time multivariate time
  series classification.
\newblock In \textit{Proceedings of the Thirty-First AAAI Conference on
  Artificial Intelligence}, 4999--5000.

\bibitem[{Wolpert(1992)}]{Wolpert1992}
Wolpert, D.~H. (1992) Stacked generalization.
\newblock \textit{Neural Networks}, \textbf{5}, 241--259.

\bibitem[{Yan and Olson(2013)}]{Yan2013}
Yan, T. and Olson, K. (2013) \textit{Improving Surveys with Paradata: Analyzing
  Paradata to Investigate Measurement Error}, chap.~4, 73--95.
\newblock John Wiley \& Sons, Ltd.

\bibitem[{Yan and Tourangeau(2008)}]{Yan2008}
Yan, T. and Tourangeau, R. (2008) Fast times and easy questions: the effects of
  age, experience and question complexity on web survey response times.
\newblock \textit{Applied Cognitive Psychology}, \textbf{22}, 51--68.

\end{thebibliography}

\end{document}